\documentclass[12pt]{article}
\usepackage{graphicx}
\usepackage{lscape}
\usepackage{citesort}
\usepackage{amssymb}
\usepackage{appendix}
\usepackage{multirow}

\begin{document}
\def\qq{\langle \bar q q \rangle}
\def\uu{\langle \bar u u \rangle}
\def\dd{\langle \bar d d \rangle}
\def\sp{\langle \bar s s \rangle}
\def\GG{\langle g_s^2 G^2 \rangle}
\def\Tr{\mbox{Tr}}
\def\figt#1#2#3{
        \begin{figure}
        $\left. \right.$
        \vspace*{-2cm}
        \begin{center}
        \includegraphics[width=10cm]{#1}
        \end{center}
        \vspace*{-0.2cm}
        \caption{#3}
        \label{#2}
        \end{figure}
	}
	
\def\figb#1#2#3{
        \begin{figure}
        $\left. \right.$
        \vspace*{-1cm}
        \begin{center}
        \includegraphics[width=10cm]{#1}
        \end{center}
        \vspace*{-0.2cm}
        \caption{#3}
        \label{#2}
        \end{figure}
                }

\def\ds{\displaystyle}
\def\beq{\begin{equation}}
\def\eeq{\end{equation}}
\def\bea{\begin{eqnarray}}
\def\eea{\end{eqnarray}}
\def\beeq{\begin{eqnarray}}
\def\eeeq{\end{eqnarray}}
\def\ve{\vert}
\def\vel{\left|}
\def\ver{\right|}
\def\nnb{\nonumber}
\def\ga{\left(}
\def\dr{\right)}
\def\aga{\left\{}
\def\adr{\right\}}
\def\lla{\left<}
\def\rra{\right>}
\def\rar{\rightarrow}
\def\lrar{\leftrightarrow}  
\def\nnb{\nonumber}
\def\la{\langle}
\def\ra{\rangle}
\def\ba{\begin{array}}
\def\ea{\end{array}}
\def\tr{\mbox{Tr}}
\def\ssp{{\Sigma^{*+}}}
\def\sso{{\Sigma^{*0}}}
\def\ssm{{\Sigma^{*-}}}
\def\xis0{{\Xi^{*0}}}
\def\xism{{\Xi^{*-}}}
\def\qs{\la \bar s s \ra}
\def\qu{\la \bar u u \ra}
\def\qd{\la \bar d d \ra}
\def\qq{\la \bar q q \ra}
\def\gGgG{\la g^2 G^2 \ra}
\def\q{\gamma_5 \not\!q}
\def\x{\gamma_5 \not\!x}
\def\g5{\gamma_5}
\def\sb{S_Q^{cf}}
\def\sd{S_d^{be}}
\def\su{S_u^{ad}}
\def\sbp{{S}_Q^{'cf}}
\def\sdp{{S}_d^{'be}}
\def\sup{{S}_u^{'ad}}
\def\ssp{{S}_s^{'??}}

\def\sig{\sigma_{\mu \nu} \gamma_5 p^\mu q^\nu}
\def\fo{f_0(\frac{s_0}{M^2})}
\def\ffi{f_1(\frac{s_0}{M^2})}
\def\fii{f_2(\frac{s_0}{M^2})}
\def\O{{\cal O}}
\def\sl{{\Sigma^0 \Lambda}}
\def\es{\!\!\! &=& \!\!\!}
\def\ap{\!\!\! &\approx& \!\!\!}
\def\ar{&+& \!\!\!}
\def\ek{&-& \!\!\!}
\def\kek{\!\!\!&-& \!\!\!}
\def\cp{&\times& \!\!\!}
\def\se{\!\!\! &\simeq& \!\!\!}
\def\eqv{&\equiv& \!\!\!}
\def\kpm{&\pm& \!\!\!}
\def\kmp{&\mp& \!\!\!}
\def\mcdot{\!\cdot\!}
\def\erar{&\rightarrow&}


\def\simlt{\stackrel{<}{{}_\sim}}
\def\simgt{\stackrel{>}{{}_\sim}}


\title{
         {\Large
                 {\bf
Determination of the mixing angle between new charmonium states
                 }
         }
      }

\author{\vspace{1cm}\\
{\small T. M. Aliev \thanks {
taliev@metu.edu.tr}~\footnote{Permanent address: Institute of
Physics, Baku, Azerbaijan.}\,\,,
M. Savc{\i} \thanks
{savci@metu.edu.tr}} \\
{\small Physics Department, Middle East Technical University,
06531 Ankara, Turkey }}

\date{}

\begin{titlepage}
\maketitle
\thispagestyle{empty}

\begin{abstract}

Using the pictures for $X(3872)$ as a mixture of charmonium and molecular
$D^{\ast 0}D^0$ states; $Y(3940)$ as a mixture of $\chi_{c0}$ and $D^\ast
D^{\prime\ast}$ states, and $Y(4260)$ as a mixture of the tetra-quark state
with charmonium states, the corresponding mixing angles are estimated within
the QCD sum rules. We obtain that our
predictions for the mixing angles of the $X(3872)$, $Y(4260)$  and $Y(3940)$
states are considerably smaller
compared to works in which mixing angles are estimated from the condition in
reproducing the the mass of these states. Our conclusion is that the
considered pictures for the $X(3872)$, $Y(4260)$  and $Y(3940)$ states are not
successful in describing these states.
\end{abstract}

~~~PACS numbers: 11.55.Hx, 12.38.Lg, 12.39.Mk, 12.39.--x
\end{titlepage}

\section{Introduction}

The analysis of the spectroscopy and decays of the heavy flavored mesons is
an essential source for obtaining useful information about the dynamics of
QCD at ``low" energies. Remarkable progress in this direction has been made
on the experimental side. Starting on from the observation of $X(3872)$
\cite{Rmes01} up to present time 23 new charmonium line states have been
discovered (see the recent review \cite{Rmes02}). All observed charmonium
states might have more complex structures compared to those predicted by the simple
quark model. These new states (referred as $XYZ$ states in the text) can be
potential candidates of exotic states, and for this reason theorists pay
great effort for understanding the dynamics of these states (for a review,
see \cite{Rmes03}). There are two attractive pictures in interpretation of
all observed states: tetra-quark,
and bound states of two mesons (meson molecules).

The theoretical approaches which have been employed in investigation of these
states are QCD sum
rules, lattice QCD, effective Lagrangian method, chiral perturbation theory,
quark model, etc. Among all approaches the QCD sum rules occupies a special
place \cite{Rmes04}, which is based on the fundamental QCD Lagrangian.
The mass and some of the strong coupling constants of $XYZ$ mesons with light
mesons are widely discussed in framework of the QCD sum rules method
\cite{Rmes03}. It is assumed in  \cite{Rmes05} that $X(3872)$ is a mixture
of charmonium and molecular states, whose interpolating current is taken as:
\bea
\label{emes01}
j_\mu \es \cos\theta j_\mu^{(2)} + \sin\theta j_\mu^{(4)}~,
\eea
where
\bea
\label{emes02}
j_\mu^{(2)} \es {1\over 6\sqrt{2}} \qq \bar{c} \gamma_\mu \gamma_5 c~, \\
\label{emes03}
j_\mu^{(4)} \es {1\over \sqrt{2}} \Big[ ( \bar{u} \gamma_5 c) (\bar{c}
\gamma_\mu u) - (\bar{u} \gamma_\mu c)(\bar{c} \gamma_5 u) \Big]~, 
\eea
and whose analysis in QCD sum rules method predicted that if the mixing
angle $\theta$ lies in the range $(9 \pm 4)^0$, it can provide good
agreement with the experimental value of the mass and decay width.
Using the QCD sum rules, the mass and the decay width of the channel $J/\Psi \omega$
for the $Y(3940)$ state is studied in \cite{Rmes06}, assuming that it is described by
the mixed scalar $\chi_{c0}$ and $D^\ast \bar{D}^\ast$ states, i.e.,
\bea
\label{emes04}
j = - {\qq \over \sqrt{2}} \cos\theta j_{\chi_{c0}} + \sin\theta j_{D^\ast
\bar{D}^\ast}~,
\eea
where $j_{\chi_{c0}} = \bar{c}c$ and $ j_{D^\ast \bar{D}^\ast} = (\bar{q}
\gamma_\mu c)(\bar{c} \gamma_\mu q)$. As a result of this study it is found
that, one can reproduce the mass and decay width of $Y(3940)$ in very good
agreement with the experimental result if the mixing angle is chosen to be
$\theta=(76\pm5)^0$.

Similar analysis is carried out in \cite{Rmes07} for the $Y(4260)$ state by
assuming that it can be described by the mixture of the tetra-quark and
charmonium currents, i.e.,
\bea
\label{emes05}
j_\mu = \cos\theta  j^{(2)}_\mu + \sin\theta j^{(4)}_\mu~,
\eea
where
\bea
\label{emes06}
j^{(4)}_\mu \es {1\over \sqrt{2}} \varepsilon_{abc} \varepsilon_{dec} \Big[
(q_a^T C\gamma_5 c_b)(\bar{q}_d \gamma_\mu\gamma_5 C\bar{c}^T) +
(q_a^T C\gamma_5 \gamma_\mu c_b)(\bar{q}_d \gamma_5 C\bar{c}^T) \Big]~, \\
\label{emes07}
j^{(2)}_\mu \es {1\over \sqrt{2}} \qq \bar{c}\gamma_\mu c~,
\eea 
and it is found that the experimental result can be reproduced by setting the
mixing angle to $\theta=(53\pm5)^0$.

Note that, the mixing studied in \cite{Rmes08} between the two- and four-quark
states is suggested for the light quark sector. This mixing is analysed with
the use of currents, and can be extended to the charm sector. The origin of
this mixing can be explained as follows: The $\bar{c}c$ state can emit a
gluon, which subsequently splits into a light quark-antiquark living like a 
molecular state during some time interval.    
In the present note we will calculate the mixing angle directly from QCD sum
rules method following the works \cite{Rmes08,Rmes09}. Let us briefly remind
the main steps of this in calculation of the mixing angle. The two physical
hadronic states are both considered to be the mixing of the two states
\bea
\label{emes08}
\eta_{H_1} =   \cos\theta \eta_{H_1^0} + \sin\theta \eta_{H_2^0}~, \nnb \\
\eta_{H_2} = - \sin\theta \eta_{H_1^0} + \cos\theta \eta_{H_2^0}~,
\eea
and then considering the correlation function
\bea
\label{emes09}
\Pi = i \int d^4x e^{ipx} \lla 0 \vel T\{\eta_{H_1}(x) \bar{\eta}_{H_2}
(0) \} \ver 0 \rra~.
\eea
According to the general strategy of the QCD sum rules method, this
correlation function is calculated in terms of hadrons on the one side;
and in terms of quarks and gluons on the other side. Using the duality ansatz
these two representations are then equated to obtain the QCD sum rules. The
correlation function from the hadronic side is calculated by saturating it
with the corresponding hadrons carrying the same quantum numbers as the
interpolating current. Obviously, the 	hadronic part of the correlation function
should be equal to zero after this procedure, since the hadronic states
given by Eq. (\ref{emes08}) are orthogonal. Using Eq. (\ref{emes08}), the
mixing angle can be calculated from Eq. (\ref{emes09}) whose expression can
be written as
\bea
\label{emes10}
\tan 2\theta = {2 \Pi^{(0)12} \over\Pi^{(0)11}-\Pi^{(0)22}}~,
\eea
where $\Pi^{(0)ij}$ are the correlation function corresponding to the
unmixed case, i.e.,
\bea          
\label{nolabel01}
\Pi^{(0)ij} = i \int d^4x e^{ipx} \lla 0 \vel T\{\eta^i(x)
\bar{\eta}^j (0) \} \ver 0 \rra~,\nnb    
\eea
where $i=1$ or $2$, and $j=1$ or 2.
In the case of scalar current,
the correlation function contains only one invariant function which we
shall denote by $\Pi^{(0)ij}$.

In the case of vector (axial-vector) current the two-point correlation
function can be written in terms of two independent invariant functions as
follows:
\bea
\label{emes11}
\Pi_{\mu\nu}^{(0)ij} (p^2) = \Pi^{(1)ij} (p^2)
\Bigg(g_{\mu\nu}-{p_\mu p_\nu \over p^2}\Bigg) +
\Pi^{(2)ij} (p^2) {p_\mu p_\nu \over p^2}~.
\eea

The invariant functions $\Pi^{(1)ij} (p^2)$ and $\Pi^{(2)ij} (p^2)$ in
Eq. (\ref{emes11}) are
associated with the spin-1 and spin-0 mesons, respectively. Since
$X(3872)$ and $Y(4260)$ mesons both have $J=1$ quantum number, 
in further discussions we shall consider only the structure
$(g_{\mu\nu} -p_\mu p_\nu / p^2)$, i.e., we shall analyse the invariant
function $\Pi^{(1)ij} (p^2)$ only.
  
Using the currents given in Eqs. (\ref{emes01}), (\ref{emes04}),
(\ref{emes07}) and their orthogonal combinations, the corresponding
correlation functions are calculated in terms of quarks and
gluons in the deep euclidean region $p^2 \small{\ll} 0$ using the operator product
expansion. The results for each current are presented in Appendix.

Having all the necessary formulas, we can now proceed
for the numerical analysis of the mixing angles. It follows from the
expressions of the mixing angles that the main input parameters involved in
the numerical calculations are the quark and gluon condensates and mass of the
quarks whose values are given as: $\qq (1~GeV) = -(246_{-19}^{+28}~MeV)^3$
\cite{Rmes10}, $\GG = 0.47~GeV^4$, $m_0^2=0.8~GeV^2$ \cite{Rmes11}. For the
mass of the $c$ quark we have used its $\overline{MS}$ value
$\bar{m}_c(\bar{m}_c)=1.28 \pm 0.03~GeV$ \cite{Rmes12}.

Beside these input parameters, the sum rules do also contain two auxiliary
parameters, namely, Borel mass parameter $M^2$ and the continuum threshold
$s_0$. The continuum threshold is correlated with the energy of the first
excited state. It is usually chosen as $\sqrt{s_0} = (m_{ground} +
0.5)~GeV$ and we look for the domain of $s_0$ which satisfies this
restriction. In this work the continuum thresholds for $X(3872)$, $Y(3940)$
and $Y(4260)$ are chosen as $\sqrt{s_0} = (4.2 \pm 0.1)~GeV$
\cite{Rmes05}, $\sqrt{s_0} = (4.4 \pm 0.1)~GeV$ \cite{Rmes06},
$\sqrt{s_0} = (4.7 \pm 0.1)~GeV$ \cite{Rmes07}, respectively, which are
obtained from the analysis of two--point function. The working region of the
Borel mass parameter $M^2$ can be obtained using the procedure. The upper
bound of $M^2$ is determined from the condition that the continuum and
higher state contributions constitute about $30\%$ of the pole contribution,
i.e.,
\bea
\label{nolabel04}
{\int_{s_0}^\infty ds \rho(s) e^{-s/M^2} \over
\int_{min}^\infty ds \rho(s) e^{-s/M^2}} < 1/3~, \nnb
\eea
where $\rho(s)$ is the spectral density which is related to the imaginary
part of the invariant function $\Pi(s)$ according to
\bea
\label{nolabel05} 
\rho(s) = {1\over \pi} \mbox{Im}\Pi(s)~. \nnb
\eea
The lower bound for $M^2$ is determined from the condition that the perturbative
contributions exceed the nonperturbative ones. From these conditions we get
the following working regions for $M^2$: $2 \le M^2 \le 4~GeV^2$ for
$X(3872)$, $Y(3940)$; and $2 \le M^2 \le 5~GeV^2$ for the
$g_{\mu\nu}-p_\mu p_\nu/p^2$ structure.

In Fig. (1) we present the dependence of the mixing angle $\theta$ on
Borel mass parameter $M^2$ at $\sqrt{s_0}= (4.4\pm 0.1)~GeV$ for the  
$g_{\mu\nu}-p_\mu p_\nu/p^2$ structure for the $X(3872)$ state. We observe from
this figure that this structure predicts for
the mixing angle $\theta\simeq (2.4\pm 0.6)^0$, which is considerably small
compared to the
value of $\theta$ obtained in \cite{Rmes05} in order to reproduce the mass
of $X(3872)$.

In Fig. (2) we present the dependence of the mixing angle $\theta$ on $M^2$ at  
fixed value of $\sqrt{s_0}=(4.4\pm 0.1)~GeV$ for the $Y(3940)$ state.
The mixing angle we obtain from this
figure is $\theta=(20\pm 2)^0$, which is about 3.5 times smaller compared to
that predicted in \cite{Rmes06}.

In Fig. (3) we present the dependence of the mixing angle $\theta$
on Borel mass parameter $M^2$ at $\sqrt{s_0}= (4.6\pm 0.1)~GeV$ for the
$g_{\mu\nu}-p_\mu p_\nu/p^2$ structure for the $Y(4260)$ state.
We see from this figure that the mixing angle has the value 
$\theta=(20\pm 3)^0$, which is approximately 2.5 times smaller than
the one predicted in \cite{Rmes07}.

In summary, in this work, based on the pictures that the $X(3872)$ is a
mixture of charmonium and $D^{\ast0} \bar{D}^0$ states, $Y(3940)$ is a
mixture of scalar $\bar{c}c$ and $D^\ast D^\ast$ molecule, and $Y(4260)$ is
a mixture of tetra-quark and charmonium states, we estimate the respective
mixing angles within the QCD sum rules method. 
It is obtained that, the mixing angles calculated in the present work
for all considered pictures are considerably smaller than the ones predicted
in \cite{Rmes05,Rmes06,Rmes07}. Therefore, in our view the considered
pictures for $X(3872)$, $Y(3940)$ and $Y(4260)$ are not successful ones in describing
these states.

\newpage

\section*{Appendix A}
\setcounter{equation}{0}
In this appendix we present the spectral densities for the $Y(4260)$,
$X(3872)$ and for the $Y(3940)$ states. Note that the spectral density is
the imaginary part of the correlation function, i.e., $\pi \rho (s) =
\mbox{Im} \Pi (q^2=s)$.

\section*{Y(4260)}

Spectral densities corresponding to $g_{\mu\nu} - p_\mu p_\nu/p^2$
structure.

\bea
\label{app02}
\rho_{11} (s) - \rho_{22} (s)\es - {1\over 3072 \pi^6}
\int_{\alpha_{min}}^{\alpha_{max}} d\alpha
\int_{\beta_{min}}^{\beta_{max}} d\beta
\Big\{12 (1-\alpha - \beta) \mu_1^2 m_q^2 (\alpha \beta \mu_1 +      
6 m_Q m_{Q^\prime} ) \nnb \\
\ek \mu_1^3 (1-\alpha - \beta) \Big[ 3 \alpha \beta  
(1+\alpha + \beta) \mu_1 - 2 (1-\alpha - \beta)^2  m_Q m_{Q^\prime}\Big] \nnb \\
\ar 16 \pi^2 m_q \Big[ 12 \alpha \beta \mu_1^2 +
m_0^2 m_Q m_{Q^\prime} + 6 (5-\alpha - \beta)
\mu_1 m_Q m_{Q^\prime} \Big] \qq \Big\} \nnb \\
\ar {\qq \over 192 \pi^4} \int_{\alpha_{min}}^{\alpha_{max}} d\alpha
\Big\{ m_q (1-\alpha) \alpha \Big[6 \mu_2^2 - m_0^2 (\mu_2 + 2s)\Big]
- 6 m_0^2 m_q  m_Q m_{Q^\prime} \nnb \\
\ek 8 \pi^2 \Big[
(1-\alpha) \alpha (m_0^2 - 7 \mu_2 + 8 s - 2 m_q^2) + 10 m_Q  
m_{Q^\prime}\Big] \Big\}~, \nnb \\ \nnb \\  
\rho_{12} (s) \es \rho_{21} (s) =
- {\qq^2\over 8 \pi^2} \int_{\alpha_{min}}^{\alpha_{max}} d\alpha
\Big[(1-\alpha) \alpha (\mu_2-s) - m_Q m_{Q^\prime} \Big]~.
\eea

\section*{X(3872)}

Spectral densities corresponding to $g_{\mu\nu} - p_\mu p_\nu/p^2$
structure.

\bea
\label{app04}
\rho_{11} (s) - \rho_{22} (s) \es {3 \mu_1 \over 4096 \pi^6} 
\int_{\alpha_{min}}^{\alpha_{max}} d\alpha
\int_{\beta_{min}}^{\beta_{max}} d\beta
\Big\{ - \alpha \beta \Big[ 1-(\alpha + \beta)^2 \Big] \mu_1^3 \nnb \\
\ar 2 (1-\alpha - \beta) \mu_1 m_q \Big[ \alpha (3 + \alpha + \beta)
\mu_1 m_{Q^\prime} + \beta (3 + \alpha + \beta) \mu_1 m_Q -    
12 m_q m_Q m_{Q^\prime} \Big] \nnb \\
\ek 8 \pi^2 \qq \Big[ m_0^2 - 2 (1 + \alpha + \beta) \mu_1 + 2 m_q^2 \Big]
( \alpha m_{Q^\prime} + \beta m_Q) \nnb \\
\ar 32 \pi^2 m_q \qq (\alpha \beta \mu_1 - 4 m_Q m_{Q^\prime}) \Big\} \nnb \\
\ar {\qq \over 512 \pi^4} \int_{\alpha_{min}}^{\alpha_{max}} d\alpha
\Big\{ 6 m_0^2 \mu_2 \Big[(1-\alpha) m_Q  + \alpha m_{Q^\prime}\Big]
- 32 \pi^2 \Big[ m_Q m_{Q^\prime} + m_q^2 \alpha (1-\alpha) \Big] \qq \nnb \\
\ar 3 m_q^2 (m_0^2 + 4 \mu_2) \Big[(1-\alpha) m_Q + \alpha m_{Q^\prime}\Big] \nnb \\
\ek 4 m_q \alpha (1-\alpha) \Big[ 3 \mu_2^2 + m_0^2 (\mu_2 - s) \Big]
- 12 m_0^2 m_q m_Q m_{Q^\prime} \nnb \\
\ar 24 \pi^2 m_q \Big[(1-\alpha) m_Q + \alpha 
m_{Q^\prime}\Big] \qq \Big\}~, \nnb \\ \nnb \\
\rho_{12} (s) \es \rho_{21} (s) =
- {\qq^2\over 96 \pi^2} \int_{\alpha_{min}}^{\alpha_{max}} d\alpha
\Big[\alpha (1-\alpha) (\mu_2-s) + m_Q m_{Q^\prime} \Big]~.
\eea

\section*{Y(3940)}

Spectral densities for the $Y(3940)$ state.

\bea
\label{app05}
\rho_{11} (s) - \rho_{22} (s) \es
{3 \mu_1 \over 512 \pi^6} \int_{\alpha_{min}}^{\alpha_{max}} d\alpha 
\int_{\beta_{min}}^{\beta_{max}} d\beta
\Big\{ -(1-\alpha - \beta) \mu_1 \Big[ \alpha \beta \mu_1^2 + 12 m_q^2
m_Q m_{Q^\prime} \nnb \\
\ek 2 \mu_1 m_q (\alpha m_{Q^\prime} + \beta m_Q ) \Big]
+ 8 \pi^2 \Big[ \mu_1  (\alpha m_{Q^\prime} + \beta m_Q ) 
- 8 m_q m_Q m_{Q^\prime} \Big] \qq \Big\} \nnb \\
\ar {\qq \over 384 \pi^4} \int_{\alpha_{min}}^{\alpha_{max}} d\alpha
\Big\{ 6 (m_0^2 + 3 \mu_2 ) m_q^2 \Big[ (1-\alpha) m_Q + \alpha
m_{Q^\prime}\Big] + 9 m_0^2 \mu_2 \Big[ (1-\alpha) m_Q + \alpha
m_{Q^\prime}\Big] \nnb \\
\ek 12 m_q \alpha (1-\alpha) \Big[ 3 \mu_2^2 + m_0^2 (2 \mu_2-s) \Big] 
-36 m_0^2 m_q m_Q m_{Q^\prime} \nnb \\
\ar 16 \pi^2 \Big[ - \alpha (1-\alpha) (14 \mu_2 - 7 s + + 9 m_q^2) + 
3 (1-\alpha) m_q m_Q + (3 \alpha m_q - 13 m_Q) m_{Q^\prime} \Big] \qq
\Big\}~, \nnb \\ \nnb \\
\rho_{12} (s) \es \rho_{21} (s) =
{\qq^2\over 4\sqrt{2} \pi^2} \int_{\alpha_{min}}^{\alpha_{max}} d\alpha  
\Big[\alpha (1-\alpha) (2 \mu_2-s) + m_Q m_{Q^\prime} \Big]~.
\eea

where,
\bea
\label{app06}
\mu_1 \es {m_Q^2 \over \alpha} +  {m_{Q^\prime}^2 \over \beta} -s~, \nnb \\
\mu_2 \es \mu (\beta \to 1-\alpha)~, \nnb \\
\beta_{min} \es {\alpha m_{Q^\prime}^2 \over s\alpha - m_Q^2}~, \nnb \\
\beta_{max} \es 1 - \alpha~, \nnb \\
\alpha_{min} \es {1\over 2s} \Big[ s + m_Q^2 - m_{Q^\prime}^2 -
\sqrt{(s+m_Q^2 - m_{Q^\prime}^2)^2 - 4 m_Q^2 s} ~ \Big]~, \nnb \\
\alpha_{max} \es {1\over 2s} \Big[ s + m_Q^2 - m_{Q^\prime}^2 +
\sqrt{(s+m_Q^2 - m_{Q^\prime}^2)^2 - 4 m_Q^2 s} ~\Big]~, \nnb
\eea

\newpage

\section*{Figure captions}
{\bf Fig. (1)} The dependence of the mixing angle $\theta$ on Borel mass
square $M^2$, at the fixed value of the continuum threshold
$\sqrt{s_0}=4.4~GeV$, for the structure $g_{\mu\nu} - p_\mu p_\nu/p^2$ for the $X(3872)$
state. \\ \\
{\bf Fig. (2)} The dependence of the mixing angle $\theta$ on Borel mass
square $M^2$, at the fixed value of the continuum threshold
$\sqrt{s_0}=4.4~GeV$, for the $Y(3940)$ state. \\ \\
{\bf Fig. (3)} The same as in Fig. (1), but at the fixed value of the
continuum threshold $\sqrt{s_0}=4.6~GeV$ for the $Y(4260)$ state.

\newpage

\begin{figure}
\vskip 3. cm
    \includegraphics{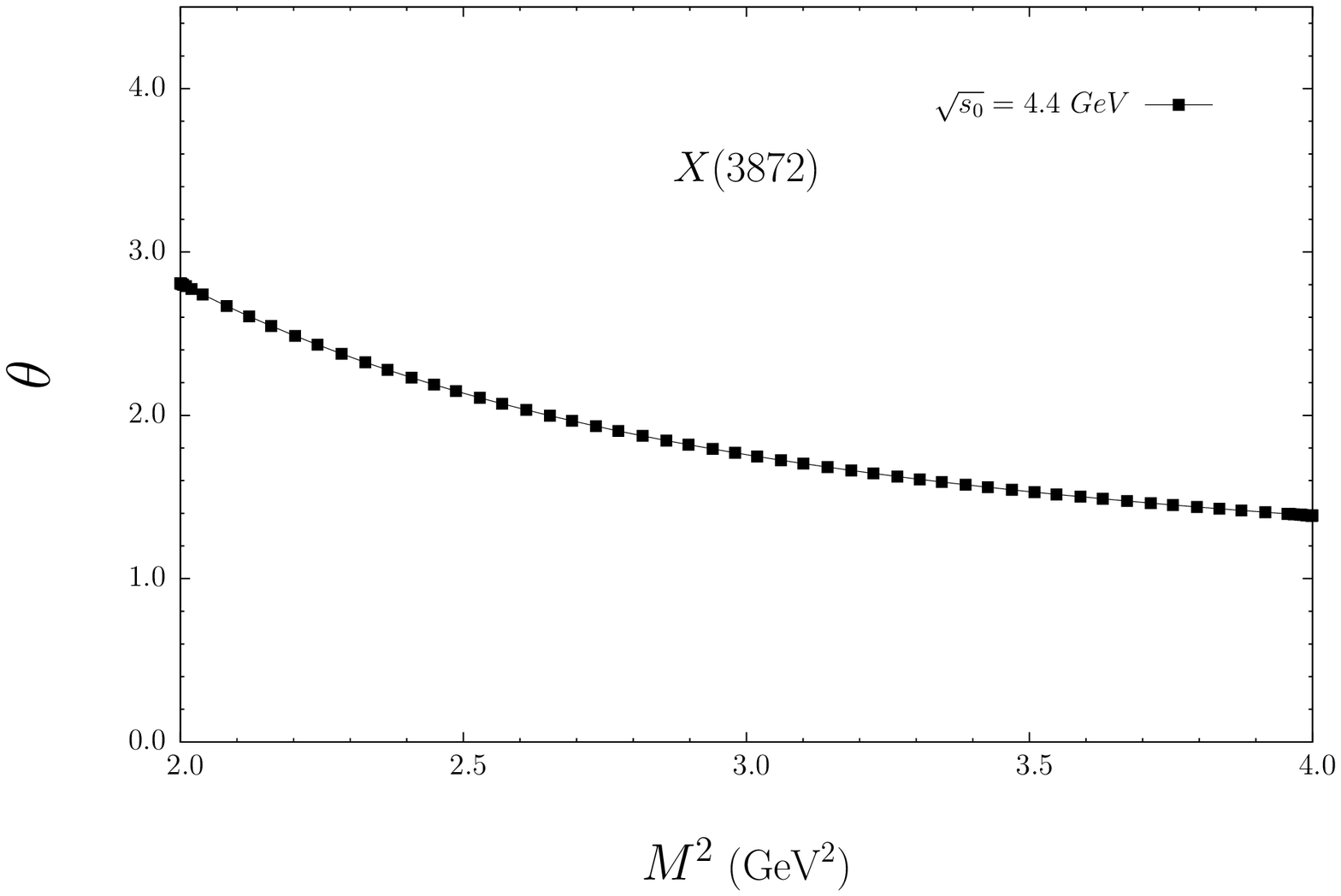}
\vskip 7.0cm
\caption{}
\end{figure}

\begin{figure}
\vskip 3. cm
    \includegraphics{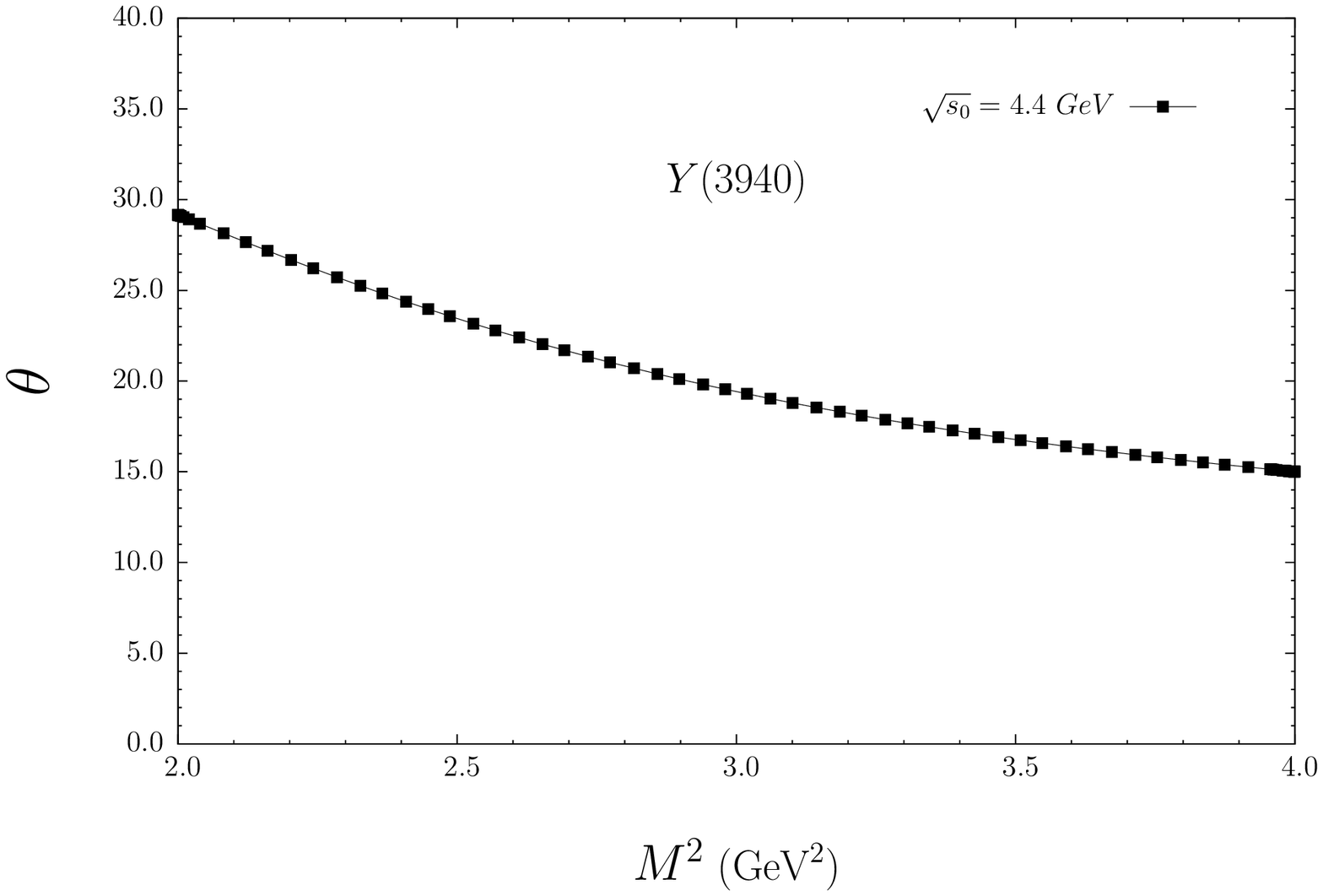}
\vskip 7.0cm
\caption{}
\end{figure}

\begin{figure}
\vskip 3. cm
    \includegraphics{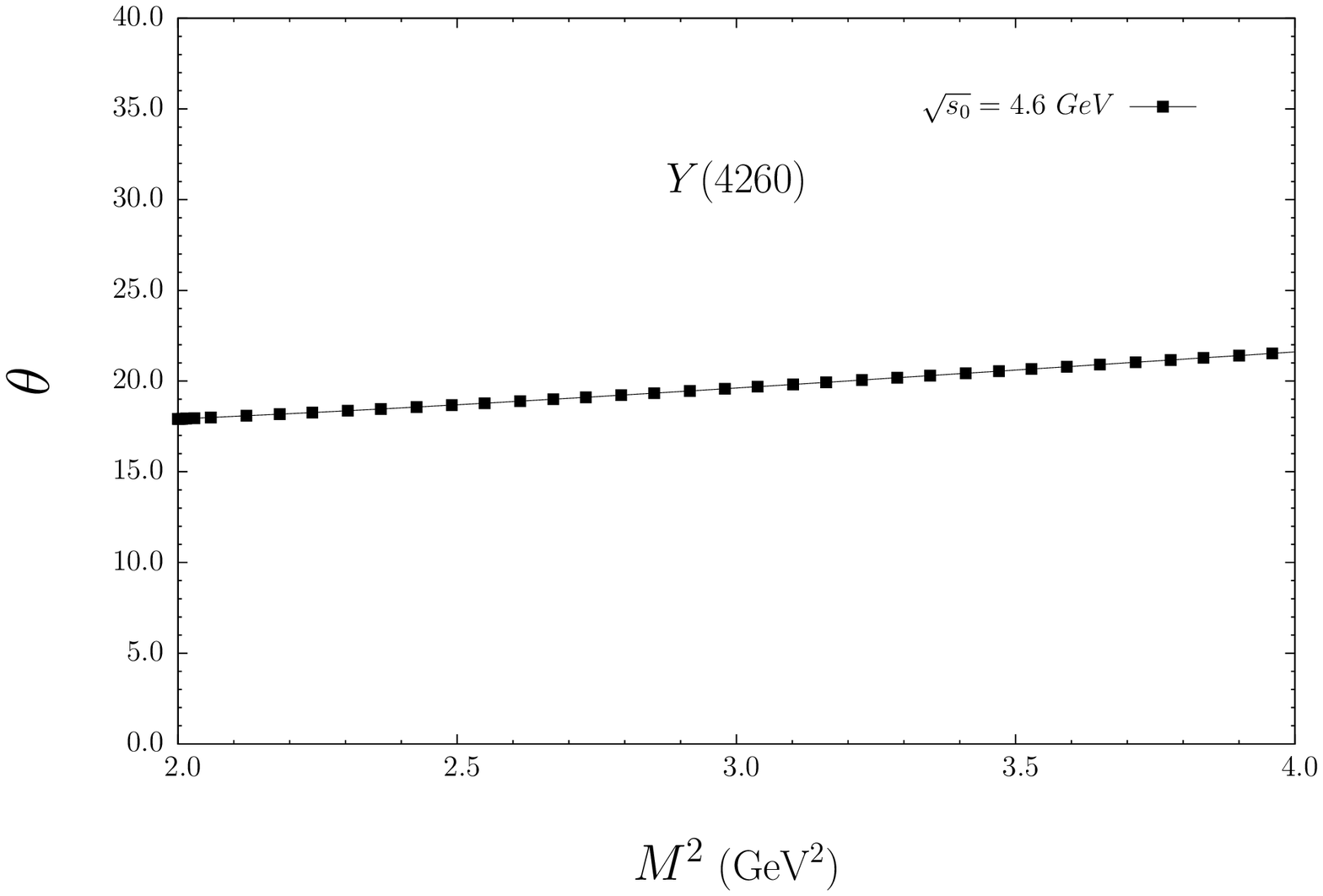}
\vskip 7.0cm
\caption{}
\end{figure}

\end{document}